\documentclass[12pt,preprint]{aastex}
 
\tighten

\def\lsim{\lower.5ex\hbox{$\; \buildrel < \over \sim \;$}}
\def\gsim{\lower.5ex\hbox{$\; \buildrel > \over \sim \;$}}
\def\lax {\ifmmode{_<\atop^{\sim}}\else{${_<\atop^{\sim}}$}\fi}
\def\gax {\ifmmode{_>\atop^{\sim}}\else{${_>\atop^{\sim}}$}\fi}

\def\gtorder{\mathrel{\raise.3ex\hbox{$>$}\mkern-14mu
\lower0.6ex\hbox{$\sim$}}}
\def\ltorder{\mathrel{\raise.3ex\hbox{$<$}\mkern-14mu
\lower0.6ex\hbox{$\sim$}}}

\slugcomment{Accepted for publication in {\it the Astrophysical Journal
Letters} 2002 August 8} 
\def\pmb#1{\setbox0=\hbox{#1}%
\kern-0.015em\copy0\kern-\wd0
\kern0.03em\copy0\kern-\wd0
\kern-0.015em\raise0.0433em\box0 }
\begin{document}

\title{On the Low  and High Frequency Correlation in Quasi-Periodic Oscillations
Among White Dwarfs, Neutron Star and Black Hole  Binaries}

\author{Lev Titarchuk \altaffilmark{1,2} and 
Kent Wood\altaffilmark{2}} 
\altaffiltext{1}{George Mason University/SCS/CEOSR, Fairfax VA
22030; lev@xip.nrl.navy.mil}
\altaffiltext{2}{Naval Research Laboratory, Washington
DC; wood@ssd5.nrl.navy.mil}

\shorttitle{Low-High Frequency Correlation}
\shortauthors{TITARCHUK and WOOD}
 
\begin{abstract}
We interpret  the correlation over five orders of magnitude between 
high frequency $\nu_{high}$ and low  frequency $\nu_{low}$ 
 in  a quasi-periodic oscillations (QPO)  found  by Psaltis, 
Belloni \& van der Klis (1999) for black hole (BH), neutron star (NS) 
systems and then extended by Mauche (2002) 
to white dwarf (WD) binaries. 
The observed correlation strongly constrains theoretical models and provides clues to understanding 
the nature of the QPO phenomena at large. 
We argue that the observed correlation is a natural consequence 
of the Keplerian disk flow adjustment to the innermost sub-Keplerian boundary conditions 
near the central object which ultimately leads to the formation of the sub-Keplerian
transition layer (TL) between the adjustment radius and
the innermost boundary (the star surface for NS and WD and the horizon for BH).
In the frameworks of the TL model $\nu_{high}$ is related to the  Keplerian  frequency 
at the outer (adjustment) radius $\nu_{\rm K}$ and $\nu_{low}$ is related to the 
magnetoacoustic oscillation frequency $\nu_{MA}$.
Using  a relation between $\nu_{MA}$ the magnetic and gas pressure  and  the  density
and  the hydrostatic equilibrium condition in the disk we  infer a linear correlation
 between $\nu_{\rm K}$ and $\nu_{MA}$, 
Identification of $\nu_{high}$, $\nu_{low}$ with 
$\nu_{\rm K}$, $\nu_{MA}$ respectively,  lead us to the determination of $H/r_{out}=1.5\times10^{-2}$ and
$\beta=0.1$ (where H is the half-width of the disk and $\beta$ is a ratio of magnetic pressure to the gas pressure). 
We estimate the magnetic field strength near the TL outer radius for BHs 
NSs and WDs.   The fact that the observed high-low frequency correlation over five orders of magnitude is valid  
for BHs, NSs, and down to WDs strongly rules out  relativistic models for QPO phenomena.
We come to the conclusion that the QPOs observations 
indicate  the  adjustment of the geometrically thin disk to sub-Keplerian motion near  the
central object. This effect is a common feature for a wide class of systems, starting
from  white dwarf binaries up to
 black hole binaries.
\end{abstract}

 
\section{Introduction}\label{sec:intro}
Quasi-periodic oscillations (QPOs) in the bright white dwarf (WD), neutron star (NS)  and 
black hole (BH)  binaries 
provide invaluable information on the accretion dynamics in the innermost parts of these systems.
The question is  what kind of  objective information 
we can extract from  observations using the fundamental principles of  fluid physics,  radiative
transfer theory and oscillatory processes? In this sense the discovery of low 20-50 Hz QPOs in the luminous NS
binaries (van der Klis et al. 1985), and discoveries of NS kilohertz QPOs by Strohmayer et al. (1996) and 
of BH hectohertz by Morgan, Remillard \& Greiner (1997) opened a new era in the study of the dynamics near
compact objects.
Psaltis, Belloni \& van der Klis (1999), hereafter PBK,  demonstrated that these NS and BH low and high frequencies 
follow a tight correlation. Their result suggests  that the same types of variability may occur in both neutron star and
black systems over 3 orders of magnitude in frequency. These two features are ``horizontal branch oscillation''
(HBO) along  with ``low frequency noise Lorentzian'', $\nu_{low}$ and the lower kHz QPO $\nu_{high}$ respectively.
Belloni, Psaltis \& van der Klis (2002), hereafter BPK,  have updated PBK's correlation adding more points from 
 Nowak (2000), Boirin et al. (2000), Homan et al. (2002), Di Salvo et al. (2001) and  Nowak et al. (2002)
data.    PBK suggest that the low and high frequencies correlate in a way that seems to depend only weakly on the
properties of the sources, such the mass, magnetic field, or possibly the presence of a hard surface in
compact object. At large scales (over 3 orders of magnitude) the PBK-BPK correlation for NS and BH systems is fitted 
by a linear function $\nu_{low}=0.081\nu_{high}$ and it also  has a fine structure at smaller frequency scales (see Fig.1). 

Recently Mauche (2002) has reported low and high frequency QPOs in the dwarf nova SS Cygni and 
VW Hyi (see  Wouldt \& Warner 2002 for  for the VW Hyi observations), and {\it has called
attention to the fact that these QPOs do extend the correlation of PBK downward in 
frequency by more than two orders of magnitude} (see Fig.1).

In fact, Titarchuk, Lapidus \& Muslimov (1998), hereafter TLM98 were first to
put forth the possibility of dynamical adjustments of a Keplerian
disk to the innermost sub-Keplerian boundary conditions to explain
most observed QPOs in bright low mass X-ray binaries (LMXBs). 
TLM98 concluded that an isothermal sub-Keplerian transition layer
between the NS surface and its last Keplerian orbit forms as a result
of this adjustment. 
The TLM98 model is a general treatment  applicable to
both NS and black hole  systems. The primary problem in both NS and BH
systems is understanding how the flow changes from pure Keplerian to the
sub-Keplerian as the radius decreases to small values. 
TLM98  suggested that the discontinuities and abrupt transitions in their
solution result from derivatives of quantities such as angular velocities
(weak shocks).
Titarchuk \& Osherovich (1999) and Titarchuk et al. (2001) interpreted the low noise  frequencies as viscous
oscillations and expected them to provide reliable estimates of radial
velocities that would be very close to the magnetoacoustic (MA) velocities.
The formulation of the MA problem and derivation of MA frequencies are presented by Titarchuk,
Bradshaw \& Wood (2001), hereafter TBW. They demonstrate that the observed correlation of the low frequency on
the low kHz frequency is perfectly described by the dependence of the inferred MA frequency on
the Keplerian frequency. 
 
In this {\it Letter}, we explain   a general correlation between low frequency and high frequency
oscillations in the TL. In \S 2, we refer to  the QPO  data obtained from  NSs and BHs by PBK and BPK, from
 the dwarf nova SS Cygni  by Mauche (2002) and from VW Hyi by Woudt \& Warner (2002). 
The formulation of the problem of low frequency oscillations and the relationship with MA oscillations
in the transition layer are described in \S 3. The derivation of the low-high frequency correlation 
 is also  present in \S 3. Our summary and conclusions follow in \S 4.

\section{Data Sources}
In Figure 1 we present data points showing  low-high frequency dependence  from BPK 
(see also PBK), Mauche (2002) and Woudt \& Warner (2002) along with a fitted line 
$\nu_{low}=0.081\nu_{high}$. The plot combines data points for BH, NS and white dwarf (WD) systems.  A reader can find 
the details of these observations and the QPO data analysis in BPK, PBK and Mauche (2002). 
{\it Each data point is strickly observational, representing two QPO frequencies that are detected at the same time in some
particular source.} 
\section{High-Low Frequency Correlation}
\subsection{The main features of the transition layer model}

TLM98 define the transition layer as a region confined between the the inner sub-Keplerian
disk boundary and the first Keplerian orbit (for the TL geometry, see Fig. 1 in TLM98 and Fig. 1 in Titarchuk, Osherovich \& Kuznetsov
1999). The radial motion in the disk is controlled by friction and
the angular momentum exchange between adjacent layers, resulting in the loss of the initial angular
momentum by accreting matter. The corresponding radial transport of the angular momentum in a disk
is described by the following equation (see e.g. Shakura \& Sunyaev 1973):
\begin{equation}
\dot M\frac{d}{dR}(\omega R^2)=2\pi\frac{d}{dR}(W_{r\varphi}R^2),
\label{eq:mot}
\end{equation}
where $\dot M$ is the accretion rate in the disk and $W_{r\varphi}$ is the component of a viscous stress
tensor that is related to the gradient of the rotational frequency $\omega=2\pi\nu$, namely,
\begin{equation}
W_{r\varphi}=-2\eta HR \frac{d\omega}{dR},
\label{eq:tens}
\end{equation}
where H is a half-thickness of a disk and $\eta$ is turbulent viscosity.
The nondimensional parameter that is essential for equation (\ref{eq:mot}) is the Reynolds number for the
accretion flow,
\begin{equation}
\gamma =\dot M/4\pi\eta H=RV_r/D,
\label{eq:rey}
\end{equation}
where $V_r$ is a characteristic velocity, and $D$ is the diffusion coefficient. D  can be defined  as $D=V_tl_t/3$ using
the turbulent velocity and  the related turbulent scale, respectively or as $D=D_M=c^2/\sigma$ for the magnetic case
where $\sigma$ is the conductivity (e.g. see details of the D-definition in Lang 1998).
Equations $\omega=\omega_0$ at $R=r_{in}$ (the inner disk radius),
$\omega=\omega_{\rm K}$ at $R=r_{out}$ (the radius where the transition layer
adjusts to the Keplerian motion), and $d\omega/dr=d\omega_{\rm K}/dr$ at
$R=r_{out}$ were assumed by TLM98 to be these boundary conditions.
Thus, the profile $\omega(R)$ and the outer radius of the transition layer
are uniquely determined by the boundary conditions   and the angular
momentum equation (\ref{eq:mot}, \ref{eq:rey})  for a given value  of Reynolds number $\gamma$.
TLM98 present the analytical solution of this TL problem (see also TO99 , Eqs. 8-9).
For the observed range of $\nu_{kQL}=400-1000$
Hz one can find using the TLM98 solution  that $\gamma$ varies from one to 5 
(see also Fig. 3 in TLM98) .

\subsection{Magnetoacoustic  oscillations in the disk transition layer} 
The frequency of the QPO associated with MA oscillations
and the correlation of the MA frequency with the Keplerian
frequency $\nu_{\rm K}$ were derived by TBW.
 The MA frequency is derived as the eigenfrequency
of the boundary-value problem resulting from a MHD treatment of the
interaction of the disk with the magnetic field.
The problem is solved for two limiting boundary conditions encompassing
realistic possibilities. The solution yields a velocity
identified as a mixture of the sound speed and the Alfv\'en velocity.
The TBW treatment does not specify
how the eigenfrequency is excited or damped. However, it makes clear that
the QPO  is a readily stimulated resonant frequency.
TBW  construct an approximate formula for the MA frequency
$\nu_{MA}$ using the asymptotic forms for the two extreme (acoustic
and magnetic) cases (see  TBW, Eqs. 12-13)
\begin{equation}
\nu_{MA}\approx [(b_s/\pi)^2\nu_{s}^2+(b_{M}/\pi)^2\nu_M^2]^{1/2},
\label{eq:ma_freq}
\end{equation}
where $b_{M}=b_s=\pi$ for the stiff boundary conditions.
$b_M$ and $b_s$ are determined by Eqs. (16-17) in TBW for the free boundary
conditions. They are approximately $b_M=\pi$ and $b_s=1$.
The frequencies $\nu_{\alpha}$,  $\nu_6=\nu_M$ ($\alpha=6$) and $\nu_0=\nu_s$ ($\alpha=0$) are determined  by the equation
\begin{equation}
\nu_{\alpha}=[(\alpha+2)/4] A^{1/2}/(r_{out}^{(\alpha+2)/2} -r_{in}^{(\alpha+2)/2}),
\end{equation}
where $A=A_s=s^2$ and  $s^2$ is the square of the sound velocity for the
pure acoustic case, and $\alpha = 6$ and
$A=A_{m}=B_{r_{out}}^2r_{out}^{\alpha}/4\pi\rho$ for the dipole magnetic case ($\rho$ is the TL mean density).
For the case of interest $r_{out}/r_{in}>2$ we approximate 
\begin{equation}
\nu_s=s/2(r_{out}-r_{in})\sim s/2r_{out} 
\end{equation}
and 
\begin{equation}
\nu_{M}=
{{2B(r_{out})}\over{(4\pi\rho)^{1/2}[r_{out}-r_{in}(r_{in}/r_{out})^ 3]}}
\sim 2V_{\rm A}(r_{out})/r_{out}.
\end{equation}
where $V_{\rm A}(r_{out}= B(r_{out})(4\pi\rho)^{-1/2}$ is the
Alfv\'en velocity at $r_{out}$. 

Now we express the MA oscillation frequencies through the gas pressure $P_g=s^2\rho$ 
and the magnetic pressure $P_{M}=V_A^2\rho/2$  using equations 
(4,6,7)
\begin{equation}
\nu_{MA}^2\approx\frac{8}{\rho r_{out}^2}(fP_g+P_{M}),
\end{equation}
where $f=1/32,~1/(32 \pi^2)$ for the stiff and free boundary conditions respectively.
\subsection{Low-high frequency correlation}
If $\beta=P_{M}/P_{g}$ is a ratio of the magnetic pressure to the gas pressure 
 we can rewrite equation (8) as follows 
\begin{equation}
\nu_{MA}^2=\frac{8P_{M}(f+\beta)}{\rho r_{out}^2\beta}.
\end{equation}
The hydrostatic equilibrium in the disk is presented by the equation
\begin{equation}
P/\rho=(P_g+P_M)/\rho=(2\pi\nu_{\rm K})^2r_{out}^2(H/r_{out})^2.
\end{equation}
Using Eq. (9) we can transform  this relation into 
\begin{equation}
\frac{(1+\beta)\nu_{MA}^2r_{out}^2}{(f+\beta)8}
=(2\pi\nu_{\rm K})^2r_{out}^2(H/r_{out})^2.
\end{equation}
which leads us  ultimately to {\it the linear relation between $\nu_{MA}$ and $\nu_{\rm K}$}
\begin{equation}
\nu_{MA}=C_{MA}\nu_{\rm K}.
\end{equation}
where $C_{MA}=2^{1/2}4\pi[(f+\beta)/(1+\beta)]^{1/2}(H/r_{out})$ is a 
proportionality coefficient which will be ``universal'' to the extent
that $\beta$ and $H/r_{out}$ remain about the same from one source to the next.
To proceed further we estimate the magnitude of $C_{MA}$ separately for NS, BH and
WD systems. 
Hydrostatic equilibrium (Eq.10) gives 
\begin{equation}
\frac{H/r_{out}}{(1+\beta)^{1/2}}=\frac{(P_g/\rho)^{1/2}}{V_{\rm K}(r_{out})}=
s/V_{\rm K}|_{r=r_{out}},
\end{equation}
where $V_{\rm K}=(GM/r)^{1/2}$ is the Keplerian velocity.
Approximate constancy within a source class  of the coefficient $C_{MA}\propto
H/r_{out}=s/V_{\rm K}$ can perhaps be understood as the thermalization of 
an excess kinetic energy according to the virial theorem. If the disk is in
equilibrium with nonradial inward motion then all of the orbital velocity at
$r_{out}$ would be needed for virial equilibrium at that radius. 
The small amount of inward motion (see Shakura \& Sunyaev 1973) creates a
disposable energy budget given $U_{excess}=m_p V_r^2/2$. 
Thus one can conclude that $(U_{excess}/U_{grav}))^{1/2}$ 
sets the maximum value of $H/r_{out}$. After that, the argument is that 
it grows to that maximum because there is little else one can do with the excess
except thermalize. The coefficient $C_{MA}$ applicable within a class is estimated
from canonical or representative values for the Keplerian frequency (essentially the lower kHz QPO
frequency $<\nu_{kQL}>$), the plasma temperature $\hat T$, all
evaluated at the transition layer outer boundary $r_{out}$ and the scaled central object mass, 
 $m=M/M_{\odot}$ (in the solar units). 
 We obtain 
\begin{equation}
C_{MA}=0.08\times k\hat T_{25}^{1/2}m_{10}^{-1/3}<\nu_{kQL,150}>^{-1/3}
(\beta/0.1)^{1/2},
\end{equation}
where $k\hat T_{25}=k\hat T/25$ keV, $m_{10}=m/10$ and $<\nu_{kQL,150}>=<\nu_{kQL}>/150$
Hz. 

Laurent and Titarchuk (1999) demonstrated that the temperature of the Compton
cooled corona $k\hat T$ is about 25 keV and 12 keV for BH and NS respectively. This assumes that
the energy released in the corona (supposedly the TL) is comparable with
the external illumination flux,  coming from the disk and from 
the star surface (in the NS case). 
Mauche (2002) and Woudt \& Warner (2002) detected the QPO frequencies from
white dwarfs (WD), SS Cyg and VW Hyi  during outbursts when the temperature of
the transition layer should be very close to the effective temperature of
 the star surface, i.e. $k\hat T=0.05-0.1$ keV.
 With assumptions $k\hat T=25,~12,~0.05$ keV, $m=10,~1.4,~ 1$, $<\nu_{kQL}>=150,~800,
 ~0.1$ Hz for BH, NS and WD respectively and $\beta=0.1$ we obtain $C_{MA}$
 values that are very close to  the observed one, $C_{MA}\approx0.08$. 
They are 0.08, 0.061 and 0.089 for BH, NS and WD respectively. 
Formula (13) also gives us estimate of the ratio $H/r_{out}$
using  $k\hat T,~m$ and $\nu_{kQL}$
\begin{equation}
H/r_{out}=0.015\times k\hat T_{25}^{1/2}m_{10}^{-1/3}<\nu_{kQL,150}>^{-1/3}.
\end{equation}
Thus $H/r_{out}\approx0.015,~0.011,~0.017$  for BH, NS and WD respectively.
\subsection{B-field estimate} 
Now we can proceed with the magnetic field strength determination in the TL using the
obtained value of $\beta=0.1$.
Because $P_M=\beta P_g$, $P_g=(kT/m_p)\rho$ and $p_M=B^2/8\pi$
we find that
\begin{equation}
B=1.14\times10^{5}k\hat T_{25}^{1/2}m_{10}^{-1/6}\tau^{1/2}<\nu_{kQL,150}>^{1/3}
~~{\rm G},
\end{equation} 
where   $\tau=(\rho/m_p)\sigma_{\rm T}r_{out}$ is the Tompson optical depth 
of the TL. Thus $B\approx1.1\times10^5,~2\times10^5,~700$ G in the TL of
BH, NS and WD systems respectively.

\section{Summary and Conclusions}
We have presented a model which treats the radial oscillations in the
transition layer surrounding a black hole,  neutron star and white dwarf.

(1) We found  that the observed high-low frequency correlation is a natural consequence 
of an adjustment of the Keplerian disk flow  to the innermost sub-Keplerian boundary conditions 
near the central object.  This ultimately leads to the formation of the sub-Keplerian
transition layer (TL) between the adjustment radius and
the innermost boundary (the star surface for NS and WD and the horizon for BH).
(2) In the framework of the TL model $\nu_{high}$ is related to the  Keplerian frequency 
at the outer (adjustment) radius $\nu_{\rm K}$ and $\nu_{low}$ is related to the 
magnetoacoustic oscillation frequency $\nu_{MA}$.
Using  a relation between $\nu_{MA}$ the magnetic and gas pressure  and  the  density
and  the hydrostatic equilibrium condition in the disk we  infer a linear correlation
 between $\nu_{\rm K}$ and $\nu_{MA}$ (see formula 12).
(3) Identification of $\nu_{high}$ and  $\nu_{low}$ with 
$\nu_{\rm K}$ and  $\nu_{MA}$ respectively,  lead us to the determination of $H/r_{out}\sim 1.5\times10^{-2}$.
(4) We present strong arguments that the ratio $\beta=P_M/P_g\approx 0.1$  in all  binaries studied.
(5) We estimate the magnetic field strength near the TL outer radius for BHs,   
NSs and WDs.  {\it (6) The fact that the observed  correlation  of high and low frequencies holds
over  five orders of magnitude shows that the QPO frequencies cannot be set by general relativistic effects,
because they are essentially non-relativistic in the white dwarf case.}  


We acknowledge Chris Mauche and Tomaso Belloni for kindly supplying us data shown in Figure 1 and  
Chris Shrader, Paul Ray  and  particularly, Chris Mauche for fruitful discussions that motivated this paper.

\clearpage
\begin{figure} 
\epsscale{0.7} 
\plotone{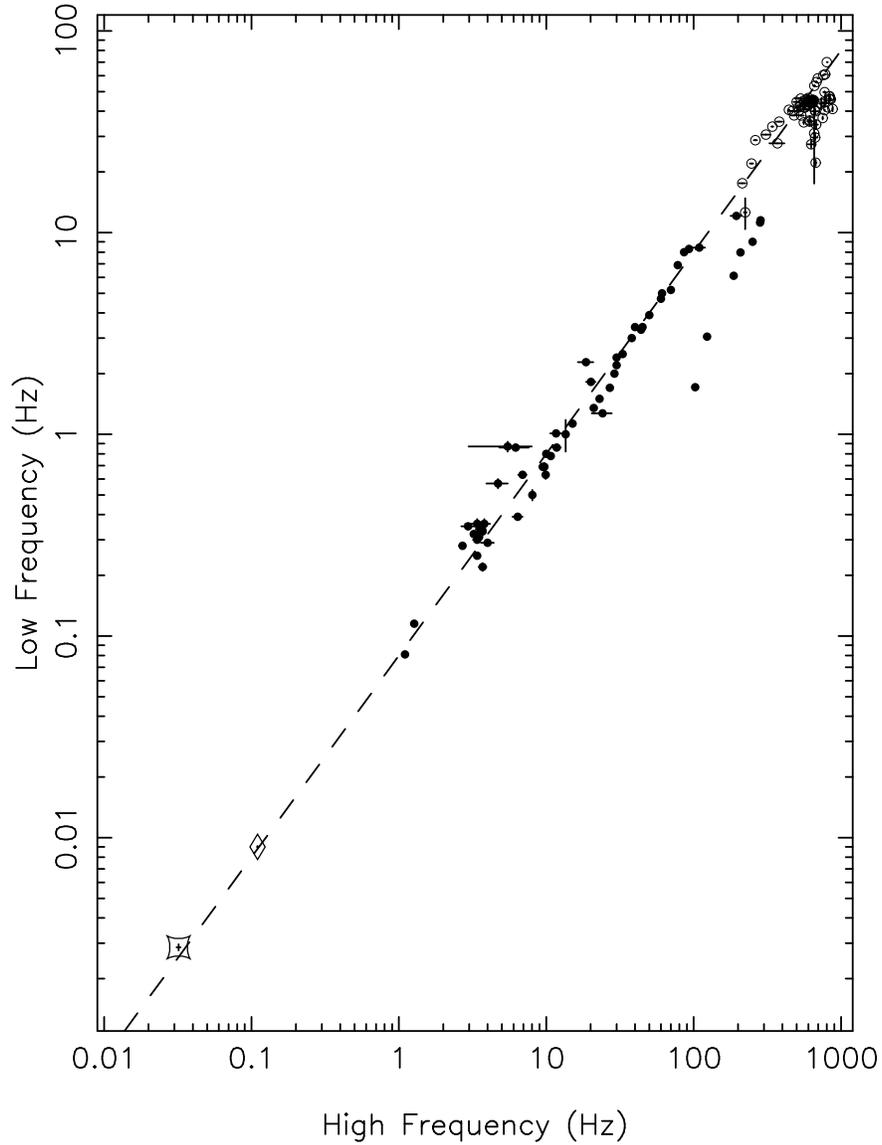} 
\caption{Correlation between frequencies of QPO and noise components white dwarfs [diamonds for SS Cyg (Mauche 2002);
squares for VW Hyi (Woudt \& Warner 2002)] , neutron star (open circles) and, 
black hole candidate (filled circles) sources.  
Neutron star and black hole data are from Belloni, Psaltis \& van der Klis (2002).
The dashed line represents the best-fit of the observed correlation.
This plot also appears in Mauche (2002).} 
\label{fig1} 
\end{figure} 


\end{document}